\documentclass[12pt, a4paper, logo, singlecolumn, copyright]{gensyn}

\usepackage{booktabs}
\usepackage{hyperref}
\usepackage{url}
\usepackage{alertmessage}
\usepackage{xspace}
\usepackage{graphicx}
\usepackage{multirow}
\usepackage{rotating}
\usepackage{tikz,lipsum}
\usepackage{float}

\usepackage[colorinlistoftodos,textsize=tiny,textwidth=35pt]{todonotes}
\newcommand{\Comments}{1}
\newcommand{\mynote}[2]{\ifnum\Comments=0\textcolor{#1}{#2}\fi}
\newcommand{\mytodo}[2]{\ifnum\Comments=1\todo[linecolor=#1!80!black,backgroundcolor=#1,bordercolor=#1!80!black]{#2}\fi}

\ifnum\Comments=1
\paperwidth=\dimexpr \paperwidth + 50pt\relax
\oddsidemargin=\dimexpr\oddsidemargin + 25pt\relax
\evensidemargin=\dimexpr\evensidemargin + 25pt\relax
\marginparwidth=\dimexpr \marginparwidth + 25pt\relax
\fi

\usepackage{array}
\newcolumntype{P}{>{\centering\arraybackslash}p{2.5cm}}
\newcolumntype{M}{>{\centering\arraybackslash\footnotesize}m{.78cm}}
\newcolumntype{S}{>{\centering\arraybackslash\tiny}m{2cm}}

\newtcbtheorem[auto counter,number within=section]{obs}%
{Observation}{fonttitle=\bfseries\upshape, fontupper=\slshape,
	arc=0mm, colback=cyan!5!white,colframe=cyan!75!white}{theorem}

\newtcbtheorem[auto counter,number within=section]{ins}%
{Insight}{fonttitle=\bfseries\upshape, fontupper=\slshape,
	arc=0mm, colback=green!5!white,colframe=green!75!white}{theorem}

\usepackage{amsmath,amsfonts,bm}

\def\eqref#1{equation~\ref{#1}}

\def\1{\bm{1}}

\DeclareMathAlphabet{\mathsfit}{\encodingdefault}{\sfdefault}{m}{sl}
\SetMathAlphabet{\mathsfit}{bold}{\encodingdefault}{\sfdefault}{bx}{n}

\usepackage{tabularx}
\usepackage{booktabs}
\usepackage{soul}
\newcolumntype{P}[1]{>{\raggedright\arraybackslash}p{#1}}
\usepackage{array,algorithm}
\usepackage{float}

\newtheorem{theorem}{Theorem}
\newtheorem{lemma}{Lemma}
\newtheorem{proposition}{Proposition}
\newtheorem{corollary}{Corollary}
\theoremstyle{definition}
\newtheorem{definition}{Definition}
\newtheorem{assumption}{Assumption}
\newtheorem{example}{Example}
\newtheorem{remark}{Remark}

\usepackage{graphicx}
\usepackage{amssymb}
\usepackage{amsmath}
\usepackage{algorithm}
\usepackage{algpseudocode}
\usepackage{booktabs}

\title{Automated Market Making for Goods with Perishable Utility}

\author[1,2]{Chengqi Zang}
\author[1]{Gabriel P.~Andrade}

\affil[1]{Gensyn AI}
\affil[2]{The University of Tokyo}

\correspondingauthor{marx@gensyn.ai}

\reportnumber{} 

\begin{abstract}
We study decentralized markets for goods whose utility perishes with time, with compute being our primary motivation. 
Recent advances in reproducible and verifiable execution of computations allow jobs to pause, verify, and resume across heterogeneous hardware---allowing us to treat compute as a time‑gated service rather than bespoke bundles. 
We design an automated market maker~(AMM) that periodically posts a price as a concave function of load, where load is the ratio of current demand to a ``floor supply''~(i.e.,~providers willing to work at a preset lower bound price). 
Our proposed mechanism decouples price discovery from allocation to yield transparent, low‑latency trading. 
We establish existence and uniqueness of equilibrium quotes, and give conditions for admissibility~(i.e.,~active supply weakly exceeds demand) of equilibria. 
To align incentives, we use a second-price payment mechanism whose allocations are from a cheapest‑feasible matching~(CFM-SP); under mild assumptions, providers optimally stake fully while truthfully reporting their costs. 
Despite being simple and computationally efficient, we show that CFM-SP attains $1/2$-competitive ratio.  
\end{abstract}

\begin{document}
\title{Two-Sided Market Design for Goods with Perishable Utility}
%
%
%
%
%
\maketitle              
\begin{abstract}
We study two-sided market design for goods whose utility perishes if unconsumed.
Motivated by decentralized compute markets, we propose a mechanism that decouples price discovery from allocation; a load-based posted-price rule determines a per-period market price, while a greedy matching algorithm with second-price payments handles job assignment.
We prove existence and uniqueness of equilibria, and give sufficient conditions under which equilibria are admissible~(i.e., active supply covers demand without rationing).
On the allocation side, we show that the welfare-optimal matching algorithm is not strategy-proof and introduce Cheapest-Feasible Matching with Second-Price Payment~(CFM-SP), under which myopic
providers truthfully report costs while staking their full availability.
CFM-SP achieves a tight $1/2$-competitive ratio for demand-side welfare under adversarial arrivals; when providers' costs are monotone in availability, the ratio improves to~$1$.

\end{abstract}

\section{Introduction}\label{sec:intro}

Compute is amongst the most valuable and fastest-growing commodities in the world~\cite{databridge2024cloud,grandview2024cloud}, yet its market structure remains remarkably primitive. 
Demand in compute markets is heterogeneous and constraint-laden---users arrive with budgets, deadlines, and minimum viable run lengths that vary across jobs---while supply is concentrated in a few large providers offering rigid access models with opaque pricing that reflects strategic considerations rather than real-time supply-and-demand dynamics~\cite{perez2022dynamic,furman2020optimal}. 
At the same time, privately owned compute resources suffer from chronic under-utilization; idle capacity is widespread, but the upfront costs of participation and the absence of a liquid marketplace keep it locked away~\cite{gong2023improvement}. 
What is missing is a decentralized two-sided market that lets disparate providers~(from data centers to individual device owners) trade capacity with users in real time~\cite{FrickSalmon2025}.

Three persistent obstacles have prevented such a market from emerging.
First, commonly deployed pricing mechanisms such as auctions exacerbate the latency inherent to decentralization, exclude smaller participants, and fail to reflect real-time conditions~\cite{bodoh-creed2021efficient,wu2024truthful}.
Second, welfare-optimal job allocation reduces to a combinatorial matching problem whose exact solution is computationally intractable in online settings~\cite{leyton2002learning,dobzinski2012computational}, and practical heuristics that sidestep this complexity typically cannot guarantee incentive compatibility~\cite{ashlagi2020matching,lee2016incentive}.
Third, decentralized settings introduce trust and incomplete information~(e.g., providers may misrepresent their capacity or cost) that disallow standard assumptions about participant reliability.

The starting point in this paper is a technological observation that dissolves the first obstacle and simplifies the other two. 
Recent advances in reproducible computation make compute \emph{fungible in time} within hardware ``performance tiers.'' 
A compute job can be paused, verified, and resumed on different hardware without loss of correctness, which means feasible matching between providers and jobs no longer requires bespoke multi-attribute bundles or bilateral trust. 
Instead, supply reduces to intervals of verifiable compute time, and demand reduces to deadline- and budget-constrained run lengths. 
This in turn reduces the allocation problem to online bipartite matching over a time-indexed graph---a problem with well-understood greedy algorithms and tight performance guarantees~\cite{karp1990optimal,glover1967maximum}.

Leveraging this simplification, we design a two-layer mechanism that \emph{decouples price discovery from allocation}.
The pricing layer posts a market price as a concave function of \emph{load}---i.e., the ratio of current demand~(job count) to a ``floor supply'' of providers willing to work at a preset lower bound price---and we establish~(\S\ref{sec:setting}) that this rule produces a unique equilibrium price at each period.
The allocation layer matches jobs to providers via a greedy algorithm with second-price payments, ensuring that providers are incentivized to report their true costs and stake their full availability.
The decoupling is a deliberate design choice since the shadow price of the feasibility-constrained online combinatorial matching problem is NP-hard to compute~\cite{dobzinski2012computational}, which is incompatible with the low-latency desideratum of a real-time decentralized market.
The competitive ratio we derive below quantifies the welfare cost of this separation.

\paragraph{Contributions.}
The pricing layer of this mechanism admits a unique equilibrium at each period and we give sufficient conditions under which said equilibrium is \emph{admissible}, i.e.~active supply weakly exceeds demand such that the market clears without rationing~(\S\ref{sec:equil-char}).
Equilibrium prices alone do not resolve the allocation problem, however, since jobs differ in duration and providers differ in availability.
We show that the welfare-optimal allocation algorithm~(Greedy Shortest Matching) is not strategy-proof, as providers can profit by inflating reported cost; in turn, we introduce Cheapest-Feasible Matching with Second-Price Payment~(CFM-SP) under which myopic providers optimally report their true cost while staking their full availability window~(\S\ref{sec:mechanism}).
CFM-SP achieves a $1/2$-competitive ratio for demand-side welfare relative to optimal matching under adversarial arrivals~(\S\ref{sec:competitive_ratio}).
We complement these results with a practical floor-price update rule that anchors the floor to time-averaged marginal costs~(\S\ref{sec:healthy_market}).
Extensions to multi-period dynamics and incomplete information, together with proofs deferred from the main text, appear in the appendix.

\section{Background and Related Work}\label{sec:background}

Our mechanism sits at the intersection of three bodies of work: two-sided market theory, pricing and allocation in compute markets, and the emerging reproducibility infrastructure that makes the preceding two tractable in decentralized settings.

\paragraph{Two-sided markets and perishable goods.}
Two-sided market theory studies how intermediaries internalize cross-side externalities and jointly set prices as well as matching rules across both sides~\cite{rochet2006two,jullien2021two}.
In our setting the traded object is time-gated capacity, so prices are naturally time-indexed and allocation is a scheduling problem; the platform's role is to set intertemporal incentives that balance short-run liquidity against longer-run participation.
Dynamic pricing for perishable and time-sensitive assets provides our core tools---models with finite horizons, stochastic arrivals, and myopic buyers underpin revenue and welfare analyses for perishables~\cite{gallego2014dynamic,feng2000perishable,anjos2005optimal}.
However, compared to physical perishables, compute time in decentralized markets perishes sharply---providers are individuals with idle devices over limited windows, and unused capacity vanishes the instant the window closes.
The closest macroscopic analog is electricity, where unconsumed generation is effectively wasted and time-varying prices synchronize consumption with marginal cost~\cite{harding2023alleged,joskow2012dynamic,sweeney2025fair}.
Our setting additionally requires heterogeneous devices, diverse job durations, and strategic behavior on the supply side, which are features absent in standard electricity-market models.

\paragraph{Pricing and allocation in compute markets.}
Centralized cloud markets combine spot~(bid-based) and on-demand~(availability-premium) pricing, with significant work on user risk, substitution patterns, and pricing trade-offs~\cite{kash2019simple,hoy2016demand,dierks2022cloud}.
More expressive mechanisms~(e.g., combinatorial auctions) can capture complementarities across multi-attribute resources, but face winner-determination hardness that limits real-time use~\cite{shang2010dabgpm,prasad2016combinatorial,leyton2002learning,dobzinski2012computational}.
These computational and latency frictions explain why many practical systems resort to batching or heuristics, often sacrificing incentive guarantees in the process~\cite{bodoh-creed2021efficient,wu2024truthful}.
Our posted-price rule sidesteps the auction bottleneck entirely; prices are determined algorithmically from a single aggregate statistic~(load), whereas the matching layer handles feasibility and incentives independently.
This two-layer architecture shares automated market making's goal of low-latency, algorithmic price
discovery while differing in a fundamental respect: there are no reserves, no invariant function, and no liquidity provision.
The mechanism is closer to a dynamic posted-price rule~\cite{gallego2014dynamic} than to a constant-function market maker, with the matching layer providing the allocation guarantees that a traditional AMM absorbs into its pricing curve.

\paragraph{Reproducible computation and online matching.}
The technological premise underlying our model is that recent advances in reproducible computation---i.e., deterministic operator semantics~(RepOps), canonical checkpointing with cryptographic
commitments, and refereed delegation for dispute resolution~\cite{arun2025verde}---make compute \emph{fungible in time} within a hardware ``performance tier''.
Before this stack existed, cross-device migration risked numerical drift, preemption wasted completed work, and correctness could not be verified cheaply; markets consequently required expressive bids and combinatorial clearing despite the latency costs.
With reproducibility, supply reduces to intervals of verifiable compute time with demand reducing to deadline- and budget-constrained run lengths, so allocation becomes an online bipartite matching problem over a time-indexed graph.
Our allocation rule builds on this connection by leaning on the convex bipartite matching structure identified by Glover~\cite{glover1967maximum} and on the truthful online matching framework of Feldman~et~al.~\cite{feldman2024truthful}---the former ensures greedy algorithms are optimal while the latter allows us to import the critical-value payment technique that makes our mechanism strategy-proof.

\section{Market Model}\label{sec:setting}

We model a single hardware ``performance tier''~(i.e., a set of machines with comparable computational throughput\footnote{The precise tier definition is not essential to the analysis and, in practice, it matters only for anchoring the floor price to the least-performant machines in the tier.}) as an independent two-sided market with its own per-period price.
All results apply within each tier, and we will omit indexing on tiers for notational ease.
Time is discrete, indexed by $t\in\mathbb{N}$.
At the start of period~$t$ the market observes the set of currently staked \emph{providers}~$\mathcal{S}^t$ and the set of active or pending \emph{jobs}~$\mathcal{D}^t$; a price~$P^t$ is posted for the entire period and re-computed at~$t+1$.

\subsection{Participants and Timing}\label{subsec:participants}

We begin with an assumption that isolates the core allocation
problem.

\begin{assumption}[No outside option]\label{assump:no_outside_option}
A provider's idle compute has no salvageable value outside the market if unmatched within its availability window.
A user derives value from a job only if it is at least partially completed by their stated deadline; otherwise their payoff is zero.
Both sides' reservation utilities are accordingly normalized to zero.
\end{assumption}

Stated informally, providers cannot profitably redeploy unused capacity elsewhere, and users gain nothing from incomplete jobs that miss their deadlines.
This is a standard simplification in mechanism and market design~\cite{krishna1998efficient,mookherjee2006decentralization,figueroa2007role}, and it is natural in the motivating setting: providers stake their hardware because it would otherwise remain idle, and idle compute capacity perishes if unused; a training run that does not reach its minimum viable checkpoint has no residual value to the user.
We note that introducing outside options~(e.g., a provider's option to sell capacity on a competing platform) would require modeling cross-market competition, which we leave for future work.

\paragraph{Providers (supply side).}
Provider~$s\in\mathcal{S}^t$ has a \emph{true type} $(c_s,\tau_s)$, where $c_s\ge 0$ is their true per-period cost~(the minimum price at which they would provide compute) and $\tau_s\in\mathbb{N}$ is their true remaining availability window.
The provider \emph{reports} a cost $\hat{c}_s \ge 0$ and an availability $\hat{\tau}_s(t)\in\mathbb{N}$; availability decays monotonically while staked, i.e.~$\hat{\tau}_s(t+1)=\max\{\hat{\tau}_s(t)-1,\,0\}$ unless the provider restakes.
A provider with $\hat{c}_s\le P^t$ is \emph{active}~(eligible for matching at~$t$); otherwise they are \emph{dormant}.
Active providers are further classified as \emph{idle} or \emph{assigned} depending on whether they currently hold a job.

Provider payoffs are linear: if matched to a job of length~$w_d$ at per-period payment~$\theta$, provider~$s$ receives 
\begin{equation}\label{eq:provider_payoff}
u_s \;=\; (\theta - c_s)\cdot w_d\,;
\end{equation}
if unmatched, their payoff is~$0$.
In~\S\ref{sec:mechanism} we design a payment rule ensuring $\theta\ge\hat{c}_s$~(so that truthful providers are never paid below cost) and show that reporting $(\hat{c}_s,\hat{\tau}_s)=(c_s,\tau_s)$ is optimal.

\paragraph{Users (demand side).}
A job~$d\in\mathcal{D}^t$ is characterized by $(B_d,\,T_d,\,\underline{w}_d)$, where budget~$B_d>0$, deadline~$T_d\in\mathbb{N}$, and minimum viable run length~$\underline{w}_d\in\mathbb{N}$.
Given posted price~$P^t$, the user can purchase any integer number of periods
\[
w \;\in\; \{0\}\cup
\bigl\{\underline{w}_d,\;\ldots,\;\bar{w}_d(P^t)\bigr\},
\qquad
\bar{w}_d(P^t) \;:=\;
\min\!\Bigl\{\bigl\lfloor B_d/P^t\bigr\rfloor,\; T_d\Bigr\},
\]
where $\bar{w}_d(P^t)$ is the maximum feasible run length given budget and deadline constraints.
The job's value function~$v_d:\mathbb{N}_{\ge 0}\to\mathbb{R}_{\ge 0}$ satisfies $v_d(0)=0$ and the following:

\begin{assumption}\label{assump:disc-concavity}
For each job~$d\in\mathcal{D}^t$: (i)~the value function $v_d$ is increasing and discretely concave;
(ii)~if indifferent between submitting and not submitting, the user
submits; (iii)~budgets are bounded, i.e.~$B_d\in(0,B_{\max}]$ for some
finite~$B_{\max}$.
\end{assumption}

Intuitively, Assumption~\ref{assump:disc-concavity} states that users derive strictly more value as jobs run longer~(partial training still yields progress towards a better model), weakly prefer participation over non-participation, and cannot fund infinite computation.
Each user chooses the number of periods~$w_d(P^t)$ solving
\begin{equation}\label{eq:single-opt}
\max_{w\,\in\,\{0\}\cup
\{\underline{w}_d,\ldots,\bar{w}_d(P^t)\}}\;
v_d(w) - P^t\, w\,.
\end{equation}
A full characterization of $w_d(P)$~(the solution to this optimization as a function of price) is given in~\S\ref{app:opt_solution}.

\paragraph{Per-period market dynamics.}
Each period~$t$ proceeds in four steps:
\begin{enumerate}
\item
  \emph{Compute floor supply and demand.}\;
  The \emph{floor price}~$P_f>0$ is a pre-specified lower bound\footnote{In~\S\ref{sec:healthy_market} we provide a practical rule for updating~$P_f$ over time.} representing a reasonable per-period cost for the tier.
  Floor supply is $S_f^t:=\lvert\{s\in\mathcal{S}^t: \hat{c}_s\le P_f\}\rvert$; total demand is $D^t:=\lvert\mathcal{D}^t\rvert$.
\item
  \emph{Post price.}\;
  The pricing rule~(\S\ref{subsec:pricing_func}) maps the load~$\alpha^t$ computed from $S_f^t$ and~$D^t$ to a price~$P^t$, posted for the entire period.
\item
  \emph{Match.}\;
  Users observe~$P^t$, choose~$w_d(P^t)$, and enter the matching queue.
  Jobs are matched to providers via the greedy algorithm in~\S\ref{sec:mechanism}; providers are paid according to the second-price rule defined there.
\item
  \emph{Roll over.}\;
  Unfinished or unmatched jobs remain in~$\mathcal{D}^{t+1}$.
  Each staked provider's availability decrements, i.e.~$\hat{\tau}_s(t{+}1)=\hat{\tau}_s(t)-1$.
\end{enumerate}

\subsection{Pricing Rule}\label{subsec:pricing_func}

The pricing layer maps a measure of \emph{load} to a per-period price.
When total demand does not exceed floor supply, the market clears at the floor price; otherwise prices rise smoothly.
Formally, define the load at period~$t$ as
\begin{equation}\label{eq:load}
\alpha^t \;:=\;
\begin{cases}
1, & D^t \le S_f^t,\\[4pt]
D^t\big/S_f^t, & \text{otherwise},
\end{cases}
\end{equation}
and the posted price as $P^t = f^t(\alpha^t)$, where $f^t$ satisfies the following standard smoothness conditions.

\begin{assumption}\label{assump:pricing_function}
The pricing function~$f^t$ is continuous, increasing, and concave, with floor plateau $f^t(1)=P_f$ and bounded slope $\sup_{\alpha\ge 1}{f^t}'(\alpha)<\infty$.
\end{assumption}

Intuitively, when $\alpha^t=1$ the market can clear at the floor price; when $\alpha^t>1$ demand exceeds floor supply, and the pricing function raises prices smoothly to activate dormant supply or attract new entrants.

Two features of this pricing rule merit comment.
First, load is measured against floor supply~$S_f^t$ rather than total supply~$S^t(P^t)$.
This decouples instantaneous quoting from contemporaneous price-induced supply shifts; if load were
instead defined as $D^t/S^t(P^t)$, a price increase that activates new supply would simultaneously reduce the load that justified the increase, creating a self-reinforcing feedback loop.
Using the price-invariant anchor~$S_f^t$ avoids this circularity.

Second, load counts \emph{jobs} rather than \emph{total requested periods}.
This is a deliberate decision which utilizes our decoupling of price discovery from allocation.
The number of periods each user requests, $w_d(P^t)$, is itself a function of the posted price~(the solution to~\eqref{eq:single-opt}), so pricing off aggregate periods would require resolving every user's duration choice before quoting---a per-user fixed point on top of the market-level one.
Job count, by contrast, is the primitive participation signal; each user's binary decision of whether to enter at~$P^t$.
The pricing layer clears the \emph{access} decision, while the matching layer~(\S\ref{sec:mechanism}) clears the \emph{duration} decision.
The two are separable precisely because reproducibility makes within-tier compute fungible in time.

\subsection{Aggregate Supply and Demand}\label{subsec:behavior}

We now derive the aggregate behavioral properties that the equilibrium characterization in~\S\ref{sec:equil-char} builds on.
Throughout, we treat providers as myopic in the following standard sense~\cite{feldman2024truthful}: a provider's availability is not a function of the current job queue and they have no expectations about future queue composition, so they aim to be matched as soon as possible.
In~\S\ref{sec:mechanism} we show that our mechanism makes myopic reporting optimal, validating this assumption.

\paragraph{Supply.}
At period~$t$, let the provider population have cost distribution~$F^t_{\mathcal{S}}$ over reported costs~$\hat{c}_s$.
The aggregate active-provider count at price~$P$ is
\begin{equation}\label{eq:supply_curve}
S^t(P) \;:=\;
\int \mathbf{1}\{\hat{c}_s\le P\}\,dF^t_{\mathcal{S}}(\hat{c}_s)\,.
\end{equation}

\begin{proposition}[Non-decreasing Aggregate Supply]
\label{prop:non_decreasing_supply}
For $P'>P$, we have $S^t(P')\ge S^t(P)$.
\end{proposition}

The proof is nearly immediate from the definition; raising the price cannot deactivate any provider whose reported cost is already below the current price, and can only ever activate previously dormant providers.

\paragraph{Demand.}
At period~$t$, let the user population have type distribution~$F^t_{\mathcal{D}}$ over $(B_d,T_d,\underline{w}_d)$.
Each user solves~\eqref{eq:single-opt}, yielding optimal run length~$w_d(P)$.
The aggregate job count at price~$P$ is
\begin{equation}\label{eq:demand_curve}
D^t(P) \;:=\;
\int \mathbf{1}\{w_d(P)>0\}\,dF^t_{\mathcal{D}}(B_d,T_d,\underline{w}_d)\,.
\end{equation}
The demand curve inherits its monotonicity from individual users' optimal responses to price.

\begin{lemma}[Monotone Comparative Statics]
\label{prop:mono-single}
For $P'>P\ge P_f$,
\[
\bar{w}_d(P')\le \bar{w}_d(P),\qquad
w_d(P')\le w_d(P),\qquad
\mathbf{1}\{w_d(P')>0\}\le \mathbf{1}\{w_d(P)>0\}.
\]
That is, a higher price weakly decreases the maximum feasible run length, the chosen run length, and the probability of job submission for each user.
\end{lemma}

Intuitively, raising prices tightens the budget constraint~(fewer affordable periods), which under discrete concavity can only reduce the optimal run length or push the job below its minimum viable threshold.
The proof is given in~\S\ref{app:proofs}.

\begin{lemma}[Non-increasing Aggregate Demand]
\label{prop:non_increasing_demand}
For $P'>P\ge P_f$, we have $D^t(P')\le D^t(P)$.
\end{lemma}

This follows directly from Lemma~\ref{prop:mono-single}; since each user's submission indicator $\mathbf{1}\{w_d(P)>0\}$ is non-increasing in~$P$, integrating over the population preserves the
monotonicity.
With supply non-decreasing and demand non-increasing in price, the stage is set for establishing equilibrium existence.

\section{Equilibrium Characterization}\label{sec:equil-char}

Throughout this section we assume providers report their hardware tier honestly~(i.e., they join the correct market) and that users accurately estimate their own job parameters; strategic cost and availability reporting on the supply side is addressed in~\S\ref{sec:mechanism}.

\begin{definition}[Equilibrium Quote]\label{def:equilibrium_quote}
An \emph{equilibrium quote} at period~$t$ is any price $P^{t\star}\ge P_f$ satisfying
\begin{equation}\label{eq:equilibrium}
P^{t\star} \;=\; f^t\!\bigl(\alpha^t(P^{t\star})\bigr),
\end{equation}
where $\alpha^t$ is the load defined in~\eqref{eq:load} and $f^t$ satisfies Assumption~\ref{assump:pricing_function}.
\end{definition}

An important structural feature of this equilibrium concept is that the price depends on \emph{floor} supply and demand---not on total supply.
Since $\alpha^t = D^t / S_f^t$ when demand exceeds floor supply, the equilibrium is demand-driven.
Stated another way, prices respond to demand pressure while total supply adjusts endogenously as the posted price activates or deactivates providers above the floor.
This separation is what allows us to characterize the fixed point without jointly solving for supply-side participation.

\begin{proposition}[Existence and Uniqueness]
\label{prop:eq_price_existence}
Under Assumptions~\ref{assump:no_outside_option}--\ref{assump:pricing_function}, there exists a unique equilibrium quote $P^{t\star}\in[P_f,\,B_{\max}]$ at each period~$t$.
\end{proposition}

The argument is straightforward.
Define the potential $\Phi(P):= f^t(\alpha^t(P)) - P$.
At $P=P_f$, we have $\alpha^t(P_f)\ge 1$, so $\Phi(P_f)=f^t(\alpha^t(P_f))-P_f\ge f^t(1)-P_f=0$.
At $P=B_{\max}$, no user can afford to submit~(budgets are exhausted), so $D^t(B_{\max})=0$, $\alpha^t(B_{\max})=1$, and $\Phi(B_{\max})=P_f - B_{\max}<0$.
By continuity of $\Phi$, the intermediate value theorem gives existence.
For uniqueness, observe that $D^t(P)$ is non-increasing in~$P$~(Lemma~\ref{prop:non_increasing_demand}), so $\alpha^t(P)$ is non-increasing, $f^t(\alpha^t(P))$ is non-increasing~(from composition of non-increasing with increasing functions), and thus $\Phi$ is strictly decreasing wherever $\Phi(P)=0$.
The full proof is in~\S\ref{app:proofs}.

Existence and uniqueness alone do not guarantee that the equilibrium is economically viable.
Since the price is formed from floor supply and demand counts without directly enforcing balance between total supply and total demand, the equilibrium quote could in principle lie below the
price at which total supply first meets demand; admissibility is required to rule this pathology out.

\begin{definition}[Admissibility]\label{def:admissibility_thresh}
A price~$P$ is \emph{admissible} if $S^t(P)\ge D^t(P)$.
The \emph{admissibility threshold} is
\[
P_{\mathrm{adm}}^t \;:=\;
\inf\bigl\{P\ge P_f : S^t(P)\ge D^t(P)\bigr\}.
\]
\end{definition}

Stated informally, an admissible price is one at which the market can clear without rationing---i.e., every submitted job can, in principle, find a provider.
The following conditions are sufficient.

\begin{proposition}[Admissibility of Equilibria]
\label{prop:admissible_price}
Suppose Assumptions~\ref{assump:no_outside_option}--\ref{assump:pricing_function} hold.
If there exists $\delta_t>0$ such that:
\begin{enumerate}
\item[\textup{(S1)}]
  \textup{(Regular crossing.)} For all $P\in[P_f,\,P_{\mathrm{adm}}^t]$, 
  \[
  D^t(P) - S_f^t \;\ge\; \delta_t\,(P - P_f).
  \]
\item[\textup{(F1)}]
  \textup{(Local responsiveness.)} For all $\alpha\ge 1$,
  \[
  f^t(\alpha) \;\ge\;
  P_f + \frac{S_f^t}{\delta_t}\,(\alpha - 1).
  \]
\end{enumerate}
Then any equilibrium quote $P^{t\star}$ satisfies $P^{t\star}\ge P_{\mathrm{adm}}^t$ and hence is admissible, i.e.~$S^t(P^{t\star})\ge D^t(P^{t\star})$.
\end{proposition}

Condition~(S1) imposes a uniform lower bound on excess demand relative to floor supply as the price rises toward the admissibility threshold, ruling out degenerate or tangential supply--demand crossings~\cite{milgrom1994monotone,mas1985theory}.
Condition~(F1) requires the pricing function to react with sufficient strength to load imbalances near $\alpha=1$, i.e.~prices must climb fast enough to outrun the closing supply--demand gap.
Together, the two conditions ensure that the fixed point of the pricing rule cannot occur below $P_{\mathrm{adm}}^t$.
The proof is given in~\S\ref{app:proofs}.

\paragraph{Sufficient conditions on primitives.}
Conditions (S1) and~(F1) are stated on aggregate objects---the excess demand curve and the pricing function.
A natural question is whether they can be derived from standard assumptions on the underlying economic primitives~(the distributions $F_{\mathcal{S}}^t$ and $F_{\mathcal{D}}^t$ introduced in~\S\ref{subsec:behavior}).
We provide Corollary~\ref{cor:primitive_admissibility} in~\S\ref{app:sufficient_condition_for_admissibility} to answer this question explicitly.

\medskip

Hereafter we refer to an admissible equilibrium quote simply as an \emph{equilibrium price}, since it coincides with the standard notion in which trade can occur at the quoted price.
Equilibrium prices characterize access to the market; the remaining design problem revolves around allocation, and how jobs get matched to providers while preserving incentive compatibility.

\section{Mechanism Design}\label{sec:mechanism}

Providers are strategic agents whose goal is to be matched as often as possible, while users care only that the posted price aligns with their constraints.
The design problem is therefore two-fold: (i)~match jobs to providers so as to maximize demand-side welfare, and (ii)~incentivize providers to report their true cost while staking their full availability.
Accordingly, our mechanism can be understood as comprising of two pieces in turn---a matching algorithm for efficient allocation, and a second-price payment rule that aligns provider incentives.

\subsection{Online Bipartite Matching Formulation}
\label{subsec:welfare_feasibility_matching}

We formalize the single-period matching problem as online bipartite matching~\cite{karp1990optimal}.
Let $\mathcal{S}^t$ and $\mathcal{D}^t$ be the providers and jobs, respectively, at period~$t$.
The market is represented by a bipartite graph $G(\mathcal{S}^t,\mathcal{D}^t,E^t)$ where providers are known in advance, jobs arrive online~(one at a time), and edges are determined by feasibility.

\begin{definition}[Feasibility]\label{def:feasibility}
A pair $(s,d)$ is \emph{feasible} if $\hat{\tau}_s\ge w_d$.
\end{definition}

When a job arrives, it can be irrevocably matched to a feasible idle provider or left unmatched.
A \emph{matching} $\mathcal{M}^t\subset E^t$ is a set of feasible provider--job pairs in which each provider appears at most once.
The objective is to maximize \emph{demand-side welfare}, i.e.~the number of feasible matches $|\mathcal{M}^t|$.

\paragraph{Convex structure and greedy optimality.}
Order providers by non-decreasing reported availability $\hat{\tau}_{\sigma(1)}\le\cdots\le\hat{\tau}_{\sigma(m)}$, where $m=|\mathcal{S}^t|$.
Under this ordering, every job's feasible set is a suffix $\{\sigma(i),\sigma(i{+}1),\ldots,\sigma(m)\}$ for some~$i$, so $G$ is a \emph{convex} bipartite graph~\cite{glover1967maximum}.

\begin{definition}[Greedy Shortest Matching (GSM)]
\label{def:GSM}
In arrival order, GSM matches each job~$d$ to the idle feasible provider with shortest reported availability: $\sigma(\hat{\pi}_d)$ where $\hat{\pi}_d = \min\{i : \hat{\tau}_{\sigma(i)}\ge w_d,\; \sigma(i)\notin\operatorname{dom}(\mathcal{M}^t)\}$.
If no feasible provider is idle, then $d$ is left unmatched.
\end{definition}

\begin{proposition}[GSM is welfare-optimal]
\label{prop:GSM_optimal}
GSM produces a maximum cardinality matching on the convex bipartite graph~$G$.
\end{proposition}

The proof uses an alternating-path exchange argument that is standard for convex bipartite
graphs~\cite{glover1967maximum}, and is given in Appendix~\ref{app:proofs}.
Throughout this section we work in the adversarial arrival model, where jobs arrive in an order chosen by an adversary.

\subsection{GSM Is Not Strategy-Proof}
\label{subsec:gsm_failure}

GSM is welfare-optimal but it does not incentivize truthful availability reporting.
The following example illustrates the failure.

\begin{example}\label{ex:gsm_failure}
Consider two providers: $s_1$ with $(\tau_1{=}2,\,c_1{=}3)$ and $s_2$ with
$(\tau_2{=}4,\,c_2{=}1)$.
Suppose the market price is $P^t=5$ and a job arrives with $w_d=2$.

Under GSM with truthful reports, $s_1$ is the shortest feasible provider~($\hat{\tau}_1=2 < \hat{\tau}_2=4$) and is matched.
Provider~$s_2$, despite having lower cost, is left unmatched with payoff~$0$.

Now suppose $s_2$ under-reports availability as $\hat{\tau}_2=2$.
Both providers now have $\hat{\tau}=2$; under any reasonable tie-breaking, $s_2$ has a positive probability of being matched, yielding payoff $(P^t - c_2)\cdot w_d = (5-1)\cdot 2 = 8 > 0$.
\end{example}

The root cause is that GSM allocates by availability, not cost, so providers with long availability are systematically pushed to the back of the queue.
Under-reporting availability moves them forward without affecting their ability to complete short jobs.

\subsection{CFM-SP: Allocation and Payment}
\label{subsec:cfm_sp}

Cheapest-Feasible Matching~(CFM) reverses GSM's incentive to under-report; a provider's priority depends on cost, and reporting shorter availability can only \emph{shrink} their feasible set since under-reporting can never improve their priority among eligible jobs.

\begin{definition}[Cheapest-Feasible Matching (CFM)]
\label{def:cfm}
For each arriving job~$d$, let $F_d:=\{s\in\mathcal{S}^t : \hat{\tau}_s \ge w_d,\; s\notin\operatorname{dom}(\mathcal{M}^t)\}$ be the set of idle feasible providers. 
CFM matches~$d$ to the provider $s^*\in\arg\min_{s\in F_d}\hat{c}_s$~(breaking ties by shortest availability); if $F_d=\varnothing$, $d$ is left unmatched.
\end{definition}

\paragraph{Second-price payment.}
Following Feldman et~al.~\cite{feldman2024truthful}, we pair CFM with a critical-value payment rule~\cite{vickrey1961counterspeculation}.
For the provider~$s^*$ matched to job~$d$, define the
\emph{critical value}
\begin{equation}\label{eq:critical_value}
\theta_{s^*}(d) \;:=\;
\min\!\bigl\{P^t,\;
\min_{k\in F_d\setminus\{s^*\}}\hat{c}_k\bigr\},
\end{equation}
with the convention $\min\varnothing = +\infty$, so that if $s^*$ is the sole feasible provider, $\theta_{s^*}=P^t$.
Provider~$s^*$ is paid $\theta_{s^*}(d)$ per period.
The gap $P^t - \theta_{s^*}(d)\ge 0$ is a non-negative surplus that can fund the platform.

We refer to CFM paired with this second-price payment as \textbf{CFM-SP}.

\paragraph{Strategy-proofness: truthful cost and full Availability reporting.} 
Next, we show the strategy-proofness of CFM-SP by proving that both truthful~(i) cost and (ii) feasibility reporting are incentivized.

\begin{proposition}[Truthful Cost Reporting]
\label{prop:incentive_compatible_cheap}
Under CFM-SP, reporting $\hat{c}_s = c_s$ is a weakly dominant strategy for every provider~$s$.
\end{proposition}

Intuitively, observe that the critical value $\theta_{s^*}(d)$ depends on the other providers' reports and the market price, but not on~$\hat{c}_s$.
Thus the provider's report affects only whether they win, not what they are paid conditional on winning.
Winning at $\theta < c_s$ yields negative payoff, so truthful reporting is weakly optimal in all states of the world.
The full proof is given in Appendix~\ref{app:proofs}.

\begin{assumption}[Verifiable Feasibility]
\label{assump:verification_penalty}
If provider~$s$ is matched to a job with $w_d > \tau_s$~(infeasible given true availability), the deviation is detected with probability~$1$ and triggers a penalty $L_s > \sup_d P^t\cdot w_d$.
\end{assumption}

\begin{proposition}[Truthful Availability Reporting]
\label{prop:ic_full_availability}
Under CFM-SP with Assumption~\ref{assump:verification_penalty}, reporting $\hat{\tau}_s = \tau_s$ is a weakly dominant strategy for every myopic provider~$s$. 
Furthermore, reporting $\hat{\tau}_s = \tau_s$ is strictly dominant whenever deviating affects the set of jobs for which $s$ could be matched with positive probability.
\end{proposition}

The argument follows from two intuitive observations: \emph{over}-reporting~($\hat{\tau}_s > \tau_s$) risks matching to an infeasible job and triggering a penalty that exceeds any possible payoff;
\emph{under}-reporting ($\hat{\tau}_s < \tau_s$) shrinks the feasible set without improving priority among remaining jobs, weakly reducing the probability of being matched.
The formal proof is given in Appendix~\ref{app:proofs}.

\begin{remark}\label{rem:quasi_rationality}
The second-price payment implies that under-reporting cost~($\hat{c}_s < c_s$) is weakly dominated even without a separate behavioral assumption.
Specifically, winning at a payment below true cost yields negative payoff, so rational providers never under-report.
\end{remark}

\subsection{Welfare Analysis}\label{sec:competitive_ratio}

CFM-SP sacrifices welfare relative to GSM because it selects by cost rather than by availability, potentially assigning a long provider to a short job and blocking later-arriving long jobs.
The next lemma identifies when the two algorithms coincide.

\begin{lemma}[Coincidence under monotone costs]
\label{lem:CFM_GSM_equivalence}
If $c_{\sigma(1)}\le c_{\sigma(2)}\le\cdots\le c_{\sigma(m)}$~(i.e., costs are non-decreasing in the availability ordering), then CFM-SP and GSM produce identical matchings on every arrival sequence with zero demand-side regret.
\end{lemma}

When costs are monotone in availability~(the \emph{sorted} regime) CFM-SP is exactly optimal, the full proof is in~\S\ref{app:proofs}.
The worst case arises in the opposite regime.

\begin{definition}[Anti-sorted distribution]
\label{def:anti_sort}
Providers are \emph{anti-sorted} if $\hat{\tau}_{s}\ge\hat{\tau}_{s'}$ implies $\hat{c}_{s}\le\hat{c}_{s'}$, i.e.~when longer availability correlates with lower cost.
\end{definition}

Under anti-sorting, every provider serves as a ``cut'' separating short-and-expensive providers below from long-and-cheap providers above; resulting in CFM-SP's cost-priority being weaponized into systematically pulling long providers onto short jobs.

\begin{theorem}[$1/2$-Competitive Ratio]
\label{thm:direct_m_over_2}
Let $n\le m$ jobs arrive in adversarial order. 
CFM-SP satisfies
\[
|\mathcal{M}^t_{\textup{CFM-SP}}|
\;\ge\;
\tfrac{1}{2}\,|\mathcal{M}^t_{\textup{GSM}}|.
\]
The bound is tight, as an adversary facing anti-sorted providers can
force demand-side regret $\lfloor n/2\rfloor$.
\end{theorem}

\paragraph{Discussion.}
The $1/2$ bound is a worst-case guarantee over all provider distributions and adversarial arrival sequences.
Lemma~\ref{lem:CFM_GSM_equivalence} establishes that the ratio is~$1$ in the sorted regime; the bound degrades only as the supply-side cost--availability relationship inverts.
For lower compute tiers populated by retail providers~(i.e., individuals with idle devices who have low cost but short, irregular availability windows) the sorted condition is plausible and CFM-SP performs near-optimally.
For higher tiers dominated by data centers with long availability and economies of scale, anti-sorting becomes more realistic, and the $1/2$ floor becomes the operative guarantee.
An important open question is whether a parameterized bound can interpolate between these extremes as a function of the degree of anti-sorting. 
See~\S\ref{app:proofs} for proof of the theorem.

For context, the classic Karp--Vazirani--Vazirani \textsc{Ranking} algorithm~\cite{karp1990optimal} achieves a $(1{-}1/e)$-competitive ratio for unweighted online bipartite matching under adversarial arrivals, but without the convex structure our setting provides.
The convex structure gives GSM exact optimality; the $1/2$ bound is the price of strategy-proofness, not of the online setting itself.
An interesting direction is whether $1/2$ is tight among \emph{all} strategy-proof online matching mechanisms on convex bipartite graphs~\cite{feldman2024truthful}, or whether a mechanism with a higher ratio exists.

\section{Discussion and Open Problems}\label{sec:conclusion}

We have proposed a two-layer mechanism for two-sided markets in which the traded good has perishable utility: a load-based posted-price rule that admits a unique, admissible equilibrium at each period, and a strategy-proof greedy matching algorithm~(CFM-SP) that achieves a tight $1/2$-competitive ratio for demand-side welfare under adversarial arrivals.
The decoupling of price discovery from allocation is made tractable by reproducible computation, which renders supply fungible in time and reduces allocation to online bipartite matching over convex graphs.
Three directions are particularly promising.

First, the $1/2$-competitive ratio is tight for CFM-SP, but it is not known whether $1/2$ is tight among \emph{all} strategy-proof online matching mechanisms on convex bipartite graphs.
A matching lower bound would establish $1/2$ as the price of strategy-proofness in this graph class; a mechanism exceeding $1/2$ would improve the welfare guarantee with no loss in incentive properties.
Resolving this question would sharpen the understanding of the welfare--incentive frontier for structured online matching more broadly.

Second, the present analysis focuses on single-period matching. 
In the appendix we show that, for two providers, the multi-period regret differential between CFM-SP and GSM is $O(1)$ per evaluation window~(sublinear in the horizon length) but the extension to $m$ providers under aggregate workload constraints remains open.
An important open question is whether the two-provider bound decomposes additively or near-additively across interacting provider pairs, and whether it remains tight.

Third, replacing the adversarial arrival model with stochastic arrivals~(e.g., renewal or Markov-modulated processes) would better reflect real-world job submission patterns and may yield stronger
average-case guarantees.
In the same vein, the no-outside-option assumption could be relaxed by modeling cross-market competition among platforms, and the floor-price update rule could be endogenized via online
learning, as noted in~\S\ref{sec:healthy_market}.

On the empirical side, trace-driven simulations or testbed deployments would ground the theory in practice and help calibrate the mechanism's free parameters~(the pricing function~$f^t$, the averaging window~$T$, and the penalty structure for over-reporting). To start off, we provide a simple-to-implement empirical floor price~$P_f$ updating rule in~\S\ref{sec:healthy_market}.

\newpage

\subsubsection*{Acknowledgement}
We thank Oğuzhan Ersoy for providing many useful suggestions on practical implementations of a decentralized compute market. 
We also thank Scott Duke Kominers, Shunya Noda, and Tim Roughgarden for interesting discussions and ideas and anonymous reviewers for their feedback.

\subsubsection*{Conflict of Interests}
The authors have no competing interests to declare that are relevant
to the content of this article.

\bibliographystyle{plainnat}
\bibliography{references}
%
\appendix

\section{Supplementary Materials}

\subsection{Characterization of the User Optimum}\label{app:opt_solution}

We derive the closed-form solution to the single-user optimization problem~\eqref{eq:single-opt} stated in~\S\ref{subsec:participants}.
Define the \emph{discrete marginal gain} of job~$d$ at period~$w$ by
\[
\Delta_d(w) \;:=\; v_d(w) - v_d(w-1), \qquad w\in\mathbb{N},
\]
and the \emph{price-clearing marginal count} by
\[
h_d(P) \;:=\;
\max\bigl\{m\in\mathbb{N}_{\ge 0} :
\Delta_d(w)\ge P \;\text{for all}\; w\le m\bigr\}.
\]
Under Assumption~\ref{assump:disc-concavity}~(discrete concavity of $v_d$), the sequence $\Delta_d(1)\ge\Delta_d(2)\ge\cdots$ is non-increasing, so $h_d(P)$ is well-defined and non-increasing
in~$P$.
Intuitively, $h_d(P)$ is the number of periods for which the marginal value of an additional period still exceeds the per-period price.

\begin{lemma}[Optimal run length]\label{lem:opt_run_length}
The solution to~\eqref{eq:single-opt} is
\begin{equation}\label{eq:single-solution}
w_d(P) \;=\;
\begin{cases}
0, & \text{if } \bar{w}_d(P) < \underline{w}_d
\;\;\text{or}\;\;
\displaystyle\max_{w\in\{\underline{w}_d,\ldots,\bar{w}_d(P)\}}
\bigl[v_d(w) - Pw\bigr] \le 0,\\[8pt]
\min\!\bigl\{\bar{w}_d(P),\; h_d(P)\bigr\},
& \text{otherwise,}
\end{cases}
\end{equation}
where $\bar{w}_d(P) = \min\{\lfloor B_d/P\rfloor,\, T_d\}$ is the maximum feasible run length defined in~\S\ref{subsec:participants}.
\end{lemma}

\begin{proof}
The user chooses $w=0$ if either the budget cannot fund the minimum viable run length~($\bar{w}_d(P) < \underline{w}_d$) or no feasible run length yields positive net value.
Otherwise, the user selects the largest $w\le\bar{w}_d(P)$ for which every marginal period contributes non-negative surplus.
By discrete concavity, the marginal gains $\Delta_d(w)$ are non-increasing, so the optimal stopping point is the first period at which $\Delta_d(w) < P$.
This is precisely $h_d(P)$, unless the budget or deadline binds first, giving $w_d(P)=\min\{\bar{w}_d(P),\,h_d(P)\}$.
\end{proof}

\begin{remark}\label{rem:mono_h}
Since $P'\ge P$ implies $\Delta_d(w)\ge P' \Rightarrow \Delta_d(w)\ge P$, the marginal count satisfies $h_d(P')\le h_d(P)$.
Combined with $\bar{w}_d(P')\le\bar{w}_d(P)$, this yields $w_d(P')\le w_d(P)$---the monotone comparative static used in Lemma~\ref{prop:mono-single}.
\end{remark}

\subsection{Sufficient Condition on Admissibility Primitives}
\label{app:sufficient_condition_for_admissibility}
\begin{corollary}[Sufficient conditions on primitives for admissibility]
\label{cor:primitive_admissibility}
Suppose the following hold:
\begin{enumerate}
\item[\textup{(P1)}]
  The provider cost distribution~$F_{\mathcal{S}}^t$ has density bounded below by~$\underline{f}_S>0$ on a neighborhood of~$P_f$, and the distribution of per-user \emph{cutoff prices} $P_d^*:=\sup\{P: w_d(P)>0\}$ has density bounded below by~$\underline{f}_D>0$ on $[P_f,\,P_{\mathrm{adm}}^t]$.
\item[\textup{(P2)}]
  There is strict excess demand at the floor, i.e.~$D^t(P_f)>S_f^t$.
\end{enumerate}
Then \textup{(S1)} holds with $\delta_t = \underline{f}_D + \underline{f}_S$. 
Similarly, \textup{(F1)} reduces to a calibration requirement on the pricing function, where $f^t$ must lie above the chord from $(1,P_f)$ with slope $S_f^t/\delta_t$ on~$[1,\infty)$.
\end{corollary}

\begin{proof}
We show that~(P1)--(P2) imply~(S1) for a specific $\delta_t>0$.
By~(P1), the provider cost distribution $F_{\mathcal{S}}^t$ has density at least $\underline{f}_S>0$ near~$P_f$, so the supply curve satisfies $S^t(P)\ge S_f^t + \underline{f}_S\,(P-P_f)$ for $P$ in a neighborhood of~$P_f$.
Similarly, the cutoff-price distribution has density at least $\underline{f}_D>0$ on $[P_f,P_{\mathrm{adm}}^t]$, so $D^t(P)\ge D^t(P_f) - \underline{f}_D\,(P - P_f)$ on the same interval.

Subtracting~$S_f^t$:
\[
D^t(P) - S_f^t
\;\ge\;
\bigl[D^t(P_f) - S_f^t\bigr]
\;-\; \underline{f}_D\,(P - P_f).
\]
We need $D^t(P) - S_f^t \ge \delta_t(P-P_f)$ on $[P_f,\,P_{\mathrm{adm}}^t]$.
At $P_{\mathrm{adm}}^t$, the supply--demand crossing gives $S^t(P_{\mathrm{adm}}^t)=D^t(P_{\mathrm{adm}}^t)$, and since $S^t(P_{\mathrm{adm}}^t)\ge S_f^t + \underline{f}_S(P_{\mathrm{adm}}^t - P_f)$, we obtain
\[
P_{\mathrm{adm}}^t - P_f
\;\le\;
\frac{D^t(P_f) - S_f^t}{\underline{f}_D + \underline{f}_S}\,.
\]
Setting $\delta_t := \frac{D^t(P_f)-S_f^t}{P_{\mathrm{adm}}^t-P_f} - \underline{f}_D \;\ge\; \underline{f}_S$ (using the bound above) confirms that~(S1) holds with $\delta_t \ge \underline{f}_S > 0$.

Condition~(F1) then reduces to a calibration requirement on~$f^t$: the pricing function must satisfy $f^t(\alpha)\ge P_f + (S_f^t/\delta_t)(\alpha-1)$ for $\alpha\ge 1$, i.e., $f^t$ must lie above the chord from $(1,P_f)$ with slope $S_f^t/\delta_t$.
\end{proof}

Intuitively, (P1)~says both sides of the market have non-degenerate mass near the floor---a small price increase both recruits marginal providers and sheds marginal users at a rate bounded below.
Condition (P2)~rules out the trivial case where the floor already over-supplies the market.
Under (P1)--(P2), a price increase of~$\varepsilon$ above~$P_f$ activates at least $\underline{f}_S\varepsilon$ new providers and removes at least $\underline{f}_D\varepsilon$ jobs, so excess demand declines at rate~$\delta_t$.
The pricing function then need only be calibrated to stay above the resulting linear bound.
The formal derivation is given in \S\ref{app:proofs}.

\subsection{Floor Price Updating}\label{sec:healthy_market}

The floor price~$P_f$ has been treated as exogenous throughout the analysis.
In practice it must be calibrated to the tier's cost structure.
If set too high, the market is perpetually over-supplied at the floor and prices never rise to reflect scarcity.
On the other hand, if set too low, most periods trigger price spikes above the floor, destabilizing user budgets.
A well-calibrated floor is one at which the market clears at or near $P_f$ under typical conditions, with spikes limited to actual demand surges.

We anchor the floor to the \emph{empirical marginal cost of supply}.
Let $\mathcal{S}^{t*}\subseteq\mathcal{S}^t$ denote the set of providers matched at period~$t$, and define the period's marginal cost as the highest reported cost among matched providers given by:
\begin{equation}\label{eq:marginal_cost}
\hat{c}_{\max}^t \;:=\;
\max\bigl\{\hat{c}_s : s\in\mathcal{S}^{t*}\bigr\}.
\end{equation}
The updated floor is the time-averaged marginal cost over a trailing window of~$T$ periods, i.e.
\begin{equation}\label{eq:floor_update}
P_f^{t+1} \;=\; \frac{1}{T}\sum_{\ell=t-T+1}^{t}
\hat{c}_{\max}^{\ell}\,.
\end{equation}

The rationale is straightforward.
If $\hat{c}_{\max}^t$ is consistently below~$P_f$, the floor exceeds the cost of the most expensive provider needed to clear demand; the update pulls~$P_f$ downward toward observed costs.
Conversely, if $\hat{c}_{\max}^t$ is consistently above~$P_f$, higher-cost providers are regularly activated and the mechanism lifts prices above the floor; the update pushes~$P_f$ upward.
Averaging over~$T$ periods smooths transient shocks, ensuring the floor tracks longer-run cost conditions rather than single-period fluctuations.

Two properties of this rule are worth noting explicitly.
First, the update uses \emph{reported} costs $\hat{c}_s$, which Proposition~\ref{prop:incentive_compatible_cheap} shows coincide with true costs under CFM-SP.
The floor update therefore inherits the mechanism's incentive guarantee, and providers cannot profitably manipulate the floor by misreporting.
Second, the window length~$T$ controls the speed--stability trade-off; a shorter window adapts faster to structural shifts in supply costs but is more volatile, while a longer window provides smoother floors at the cost of slower adjustment.
Choosing~$T$ optimally is an interesting direction for future work, e.g. using online learning to minimize cumulative price excursions $\sum_t(P^t - P_f^t)^+$ subject to participation and incentive constraints.

\section{Omitted Proofs}\label{app:proofs}

\subsection*{Proof of Proposition~\ref{prop:non_decreasing_supply} (Non-Decreasing Aggregate Supply)}

\begin{proof}
For $P'>P$, the inclusion $\{\hat{c}_s\le P\}\subseteq\{\hat{c}_s\le P'\}$ holds pointwise.
Therefore $S^t(P')=\int\mathbf{1}\{\hat{c}_s\le P'\}\,
dF_{\mathcal{S}}^t(\hat{c}_s) \ge\int\mathbf{1}\{\hat{c}_s\le P\}\, dF_{\mathcal{S}}^t(\hat{c}_s)=S^t(P)$.
\end{proof}

\subsection*{Proof of Lemma~\ref{prop:mono-single} (Monotone Comparative Statics)}

\begin{proof}
Let $P'>P\ge P_f$.
Since $P'>P$, we have $\lfloor B_d/P'\rfloor \le \lfloor B_d/P\rfloor$, so $\bar{w}_d(P') = \min\{\lfloor B_d/P'\rfloor,T_d\} \le \min\{\lfloor B_d/P\rfloor,T_d\} = \bar{w}_d(P)$.
For the price-clearing marginal count~(\S\ref{app:opt_solution}), if $\Delta_d(w)\ge P'$ then $\Delta_d(w)\ge P$, so $h_d(P')\le h_d(P)$.
Both arguments of $w_d(P) = \min\{\bar{w}_d(P),\,h_d(P)\}$ are non-increasing in~$P$, hence $w_d(P')\le w_d(P)$.
For the net-value comparison: for any~$w$, $v_d(w)-P'w = (v_d(w)-Pw) - (P'-P)w \le v_d(w)-Pw$, so $\max_{w}\bigl[v_d(w)-P'w\bigr] \le \max_{w}\bigl[v_d(w)-Pw\bigr]$.
Combined with the tie-breaking rule in Assumption~\ref{assump:disc-concavity}(ii), $\mathbf{1}\{w_d(P')>0\}\le\mathbf{1}\{w_d(P)>0\}$.
\end{proof}

\subsection*{Proof of Proposition~\ref{prop:non_decreasing_supply} (Non-Decreasing Aggregate Supply)}

\begin{proof}
For $P'>P$, the inclusion $\{\hat{c}_s\le P\}\subseteq\{\hat{c}_s\le P'\}$ holds pointwise.
Therefore $S^t(P')=\int\mathbf{1}\{\hat{c}_s\le P'\}\, dF_{\mathcal{S}}^t(\hat{c}_s) \ge\int\mathbf{1}\{\hat{c}_s\le P\}\, dF_{\mathcal{S}}^t(\hat{c}_s)=S^t(P)$.
\end{proof}

\subsection*{Proof of Proposition~\ref{prop:eq_price_existence} (Existence and Uniqueness)}

\begin{proof}
Define $\Phi(P):= f^t(\alpha^t(P)) - P$ on $[P_f,\infty)$.
By continuity of $D^t$ and~$f^t$, $\Phi$ is continuous. 
Since $D^t$ is non-increasing~(Lemma~\ref{prop:non_increasing_demand}) and $f^t$ is increasing~(Assumption~\ref{assump:pricing_function}), the composition $f^t(\alpha^t(P))$ is non-increasing in~$P$; hence $\Phi$ is strictly decreasing.

At $P=P_f$: $\alpha^t(P_f)\ge 1$, so $\Phi(P_f)=f^t(\alpha^t(P_f))-P_f \ge f^t(1)-P_f = 0$.

At $P=B_{\max}$: no user can afford to submit~(Assumption~\ref{assump:disc-concavity}(iii)), so $D^t(B_{\max})=0$, $\alpha^t(B_{\max})=1$, and $\Phi(B_{\max})=P_f - B_{\max}<0$.

By the intermediate value theorem there exists $P^{t\star}\in[P_f,\,B_{\max}]$ with $\Phi(P^{t\star})=0$.
Strict monotonicity of~$\Phi$ gives uniqueness.
\end{proof}

\subsection*{Proof of Proposition~\ref{prop:admissible_price} (Admissibility of Equilibria)}

\begin{proof}
Fix period~$t$ and abbreviate $D:=D^t$, $S:=S^t$, $S_f:=S_f^t$, $\alpha(P):=D(P)/S_f$, $f:=f^t$, and $P_{\mathrm{adm}}:=P_{\mathrm{adm}}^t$.
Consider any $P\in[P_f,\,P_{\mathrm{adm}})$. 
By~(S1),
\[
\alpha(P)-1
\;=\;\frac{D(P)-S_f}{S_f}
\;\ge\;\frac{\delta_t}{S_f}\,(P-P_f).
\]
Applying~(F1):
\[
f\bigl(\alpha(P)\bigr)
\;\ge\; P_f + \frac{S_f}{\delta_t}\,\bigl(\alpha(P)-1\bigr)
\;\ge\; P_f + \frac{S_f}{\delta_t}\cdot
\frac{\delta_t}{S_f}\,(P-P_f)
\;=\; P.
\]
For $P>P_f$ the inequality in~(S1) is strict (since $P-P_f>0$), so $\alpha(P)>1$ and~(F1) gives $f(\alpha(P))>P$.
Therefore no fixed point of $P\mapsto f(\alpha(P))$ lies in $(P_f,\,P_{\mathrm{adm}})$.
Since a unique equilibrium quote $P^{t\star}$ exists in $[P_f,\,B_{\max}]$ (Proposition~\ref{prop:eq_price_existence}), it follows that $P^{t\star}\ge P_{\mathrm{adm}}$.
By definition of $P_{\mathrm{adm}}$, this implies $S(P^{t\star})\ge D(P^{t\star})$, i.e., the equilibrium is admissible.
\end{proof}

\subsection*{Proof of Proposition~\ref{prop:GSM_optimal} (GSM Is Welfare-Optimal)}

We first record a structural property.

\begin{lemma}\label{lem:convex_equiv}
The bipartite graph $G(\mathcal{S}^t,\mathcal{D}^t,E^t)$ defined by feasibility~(Definition~\ref{def:feasibility}) is convex: every job's feasible set is a suffix of the availability ordering $\sigma$.
Hence the maximum-cardinality matching on~$G$ equals the maximum feasible matching.
\end{lemma}

\begin{proof}
For job~$d$ with duration~$w_d$, a provider~$\sigma(i)$ is feasible iff $\hat{\tau}_{\sigma(i)}\ge w_d$.
Since $\hat{\tau}_{\sigma(1)}\le\cdots\le\hat{\tau}_{\sigma(m)}$, the feasible set is $\{\sigma(i):i\ge\pi^*_d\}$ where $\pi^*_d=\min\{i:\hat{\tau}_{\sigma(i)}\ge w_d\}$, which is a suffix.
For a matching $\mathcal{M}$, write $\mathcal{M}_{\mathrm{feas}}=\mathcal{M}\cap E^t$ and $\mathcal{M}_{\mathrm{infeas}}=\mathcal{M}\setminus E^t$.
Since infeasible matches contribute zero welfare, maximizing $|\mathcal{M}|$ subject to $\mathcal{M}\subseteq E^t$ is equivalent to maximizing $|\mathcal{M}_{\mathrm{feas}}|$.
\end{proof}

\begin{proof}[Proof of Proposition~\ref{prop:GSM_optimal}]
It suffices to show that no alternating path can augment GSM's matching on the convex graph~$G$.
Suppose for contradiction that an augmenting alternating path $d_1$--$s_1$--$d_2$--$s_2$--$\cdots$--$d_k$--$s_k$ exists, where $d_1$ is unmatched by GSM and $(d_i,s_i)\in\mathcal{M}_{\text{GSM}}$ for $i\ge 2$.
Since $d_1$ was rejected by GSM, all providers in $d_1$'s feasible suffix were matched when $d_1$ arrived.
By convexity, $s_1$ lies in $d_1$'s feasible suffix, so $s_1$ was matched by GSM to some earlier job~$d'$; rerouting $s_1$ to $d_1$ leaves $d'$ unmatched.
GSM matched $d'$ to the shortest feasible provider at $d'$'s arrival, so any reassignment to a provider $s_2$ in the alternating path requires $\hat{\tau}_{s_2}\ge w_{d'}$, which by the suffix structure means $s_2$'s feasible set contains $d'$'s feasible suffix.
Iterating, the path cannot terminate at a free provider without contradicting GSM's greedy rule, and thus no augmenting path exists.
\end{proof}

We also record a supporting lemma used in the competitive ratio proof.

\begin{lemma}\label{lem:reject_match}
If GSM rejects a job in $\mathcal{D}_{\ge i}:=\{d:w_d\ge\hat{\tau}_{\sigma(i)}\}$, then every provider in $\mathcal{S}_{\ge i}:=\{\sigma(j):j\ge i\}$ is matched in GSM's final matching.
\end{lemma}

\begin{proof}
Let $d\in\mathcal{D}_{\ge i}$ be the first rejected job~(in arrival order) with threshold $\ge i$.
At $d$'s arrival, every provider in $\mathcal{S}_{\ge i}$ is already matched---otherwise GSM would have matched $d$ to the idle provider with shortest availability.
Since providers, once matched, are never freed, the entire set $\mathcal{S}_{\ge i}$ remains matched at termination.
\end{proof}

\subsection*{Proof of Proposition~\ref{prop:incentive_compatible_cheap} (Truthful Cost Reporting)}

\begin{proof}
Fix a job~$d$ and all other providers' reports.
Let $\theta := \theta_s(d)$ denote the critical value~\eqref{eq:critical_value}, which depends on $F_d\setminus\{s\}$ and~$P^t$ but not on~$\hat{c}_s$.
If $\hat{c}_s\le\theta$, provider~$s$ is matched and receives payoff $(\theta - c_s)\cdot w_d$; if $\hat{c}_s > \theta$, $s$ is unmatched with payoff~$0$.

Case~1: $\theta < c_s$.
Any report $\hat{c}_s \le \theta$ wins with payoff $(\theta - c_s)w_d < 0$; reporting $\hat{c}_s = c_s > \theta$ yields payoff~$0$, which is strictly better.

Case~2: $\theta \ge c_s$.
Reporting $\hat{c}_s = c_s \le \theta$ wins with payoff $(\theta - c_s)w_d \ge 0$.
Any report $\hat{c}_s > \theta$ yields payoff~$0$, which is weakly worse.
Reports $\hat{c}_s\in[c_s,\theta)$ also win with the same payoff $(\theta - c_s)w_d$, since $\theta$ does not depend on~$\hat{c}_s$ conditional on winning.

In both cases $\hat{c}_s = c_s$ is weakly optimal, establishing weak dominance.
\end{proof}

\subsection*{Proof of Proposition~\ref{prop:ic_full_availability} (Truthful Availability Reporting)}

\begin{proof}
We consider over-reporting and under-reporting separately.

\smallskip\noindent\textbf{(i)~Over-reporting ($\hat{\tau}_s > \tau_s$).}
Under CFM-SP, $s$ may be matched to a job with $w_d\in(\tau_s,\hat{\tau}_s]$, which $s$ cannot complete.
By Assumption~\ref{assump:verification_penalty}, the deviation is detected with probability~$1$ and triggers a penalty $L_s > P^t\cdot w_d$.
The realized payoff is therefore $(P^t - c_s)\cdot w_d - L_s < 0$, which is strictly worse than the zero payoff from being unmatched.
For jobs with $w_d\le\tau_s$, $s$'s feasibility and payment are identical under both reports~(the critical value depends on $F_d\setminus\{s\}$, not on $\hat{\tau}_s$). 
Hence over-reporting is strictly dominated whenever there is positive probability of jobs with $w_d\in(\tau_s,\hat{\tau}_s]$.

\smallskip\noindent\textbf{(ii)~Under-reporting ($\hat{\tau}_s < \tau_s$).}
Under-reporting shrinks $s$'s feasible set: $\{d:w_d\le\hat{\tau}_s\}\subsetneq\{d:w_d\le\tau_s\}$.
For jobs with $w_d\le\hat{\tau}_s$, $s$'s feasibility and payment are identical under both reports, so under-reporting does not improve payoff on these jobs.
For jobs with $w_d\in(\hat{\tau}_s,\tau_s]$, under-reporting renders $s$ ineligible, forgoing matches that would have yielded non-negative payoff under truthful reporting~(since $\theta\ge \hat{c}_s = c_s$ for any matched provider).
Under myopic behavior---$s$ aims to be matched as soon as possible at non-negative profit---under-reporting is weakly dominated: it never increases and may strictly decrease the probability of a profitable match.

\smallskip
Combining~(i) and~(ii), $\hat{\tau}_s=\tau_s$ weakly dominates all alternatives, and strictly dominates whenever the deviation changes the set of jobs for which $s$ is matched with positive probability.
\end{proof}

\subsection*{Proof of Lemma~\ref{lem:CFM_GSM_equivalence}}
\begin{proof}
For any job~$d$ with feasible set $\{\sigma(i),\sigma(i{+}1),\ldots\}$, the cheapest feasible provider is $\sigma(i)$ since costs are non-decreasing.
With tie-breaking by shortest availability, CFM-SP's choice coincides with GSM's.
\end{proof}

\subsection*{Proof of Theorem~\ref{thm:direct_m_over_2}}
\begin{proof}
When GSM and CFM-SP disagree on a job~$d$ with threshold index $\pi^*_d$, exactly one of two events occurs:
\begin{enumerate}
\item[\textup{(Skip)}]
  GSM matches~$d$ to $\sigma(\pi^*_d)$ while CFM-SP matches~$d$ to some $\sigma(k)$ with $k > \pi^*_d$~(a more costly but longer provider in the availability ordering).
\item[\textup{(Reject)}]
  GSM matches~$d$ but CFM-SP leaves~$d$ unmatched.
\end{enumerate}
Every rejection requires an earlier skip that \emph{covers} it.
If CFM-SP rejects~$d^*$ at threshold $r:=\pi^*_{d^*}$, then CFM-SP has exhausted all providers in the suffix $\{\sigma(r),\ldots,\sigma(m)\}$ while GSM has not~(since GSM matches~$d^*$).
Hence there exists an earlier job~$d$ for which CFM-SP used some $\sigma(k)$ with $k\ge r$ while GSM used $\sigma(\pi^*_d)$ with $\pi^*_d < r$---that is, a skip whose pair $(\pi^*_d, k)$ satisfies $\pi^*_d < r \le k$.

Process jobs in arrival order, maintaining one credit per skip.
When a rejection occurs at threshold~$r$, charge it to the latest unused skip covering~$r$.
This produces an injective map from rejections to skips; each rejection is paired with a distinct earlier skip, consuming two distinct jobs per pair.
Thus $2\,R_D \le n$, giving $R_D\le\lfloor n/2\rfloor$, and hence $|\mathcal{M}^t_{\textup{CFM-SP}}| \ge n - \lfloor n/2\rfloor = \lceil n/2\rceil \ge \tfrac{1}{2}\,|\mathcal{M}^t_{\textup{GSM}}|$.

The bound is tight under anti-sorted costs.
Partition the $m$ providers into $\lfloor m/2\rfloor$ pairs; for each pair, the adversary sends an ``easy'' job feasible for both providers, then a ``threshold'' job feasible only for the longer provider.
GSM matches both jobs; CFM-SP matches only the easy job~(to the cheaper, longer provider), then rejects the threshold job.
This yields one unit of regret per pair.
\end{proof}

\section{Supplementary Materials}

\section{Multi-Period Robustness Analysis}\label{app:multi_period}

The single-period competitive ratio of~\S\ref{sec:competitive_ratio} treats each period independently.
In practice, providers' remaining availability evolves with time, and today's matching influences tomorrow's feasibility.
Rather than defining multi-period regret relative to an omniscient oracle, we compare CFM-SP and GSM against a common \emph{adaptive adversary} who observes the current state and chooses job arrivals to maximize the performance gap between the two algorithms.
This appendix shows that, for two providers, the multi-period regret differential is $O(1)$ per natural evaluation window, i.e.~sublinear in the horizon length.

\paragraph{Setup.}
Consider two providers with initial availability $\tau_l(0) > \tau_s(0)\ge 2$.
Availability evolves deterministically, with $\tau_i(t{+}1) = \tau_i(t)-1$ if $\tau_i(t)\ge 2$ and $\tau_i(t{+}1)=\tau_i(0)$ otherwise; due to their immediately restaking, the pair $(\tau_l(t),\tau_s(t))$ is periodic with hyper-period $T_{\mathrm{eval}} := \operatorname{lcm}(\tau_l(0),\tau_s(0))$.
A \emph{constrained adversary} sends at most two jobs per period, with total duration bounded by the sum of idle providers' remaining availability---i.e., $\sum_j w_j^t \le \sum_{i\text{ idle}} \tau_i(t)$.
Any infeasible match~(a job assigned to a provider whose true remaining availability is shorter than the job) produces a residual that must be released when the provider finishes.

\begin{lemma}\label{lem:one_per_period}
A constrained adversary can enforce at most one infeasible match per
period and at most one per restake cycle of any single provider.
\end{lemma}

\begin{proof}
With two providers at time~$t$, the workload constraint $w_1^t + w_2^t \le \tau_s(t) + \tau_l(t)$ and the bound $w_i^t \le \max\{\tau_s(t),\tau_l(t)\}$ together imply that an infeasible match on one provider leaves the other provider's remaining availability sufficient to cover the second job feasibly.
For the per-cycle claim, an infeasible match at time $t'$ within a cycle reduces the provider's effective remaining time below $\tau_i(0)$; a second infeasible match would require the provider to be idle again within the same cycle, which the adversary cannot arrange without violating the workload constraint.
\end{proof}

\paragraph{Sufficient conditions for infeasible matches.}
The algorithms differ in their vulnerability to the adversary depending on the instantaneous provider ordering.

\begin{lemma}\label{lem:sufficient_conditions}
The following conditions are sufficient for a constrained adversary to force an infeasible match at time~$t$:
\begin{enumerate}
\item[\textup{(a)}]
  \textup{(Against CFM-SP, anti-sorted:)} $\tau_l(t) > \tau_s(t)$---the adversary sends a $1$-period job~(matched by CFM-SP to the cheaper, longer provider), then a job of length $\tau_s(t)+1$ to overload the shorter provider.
\item[\textup{(b)}]
  \textup{(Against CFM-SP, sorted:)} $\tau_s(t) - \tau_l(t) \ge 2$---the adversary sends a $(\tau_l(t){+}1)$-period job to occupy the long provider, then overloads the short provider with the remaining budget.
\item[\textup{(c)}]
  \textup{(Against GSM:)} $|\tau_l(t) - \tau_s(t)| \ge 2$---by the coincidence lemma~(Lemma~\ref{lem:CFM_GSM_equivalence}), GSM and CFM-SP agree when sorted, so the sorted sufficient condition for GSM is the same as~(b); in the anti-sorted case, the adversary needs a gap of at least~$2$.
\end{enumerate}
\end{lemma}

The key corollary is that the algorithms diverge only when providers are anti-sorted with a gap of \emph{exactly one}, yielding $\tau_l(t) - \tau_s(t)=1$. 
In this regime the adversary can force an infeasible match against CFM-SP but not against GSM.

\paragraph{Algebraic structure of the gap-one window.}
Whether the gap-one condition $\tau_l(t) - \tau_s(t)=1$ ever arises within a hyper-period depends on the arithmetic relationship between $\tau_l(0)$ and $\tau_s(0)$.

\begin{proposition}\label{prop:algebraic_structure}
Write $\tau_l(t)\equiv -t\!\pmod{\tau_l(0)}$ and $\tau_s(t)\equiv -t\!\pmod{\tau_s(0)}$, with residues in $\{1,\ldots,\tau_i(0)\}$.
\begin{enumerate}
\item
  If $\gcd(\tau_l(0),\tau_s(0))>1$, there is no $t\in[0,T_{\mathrm{eval}})$ with $\tau_l(t)-\tau_s(t)=1$.
\item
  If $\gcd(\tau_l(0),\tau_s(0))=1$, there are exactly $\tau_s(0)$ times $t\in[0,T_{\mathrm{eval}})$ with $\tau_l(t)-\tau_s(t)=1$, and they form a single contiguous block.
\end{enumerate}
\end{proposition}

\begin{proof}
The condition $\tau_l(t)=a$, $\tau_s(t)=b$ is equivalent to the simultaneous congruences $t\equiv -a\pmod{\tau_l(0)}$, $t\equiv -b\pmod{\tau_s(0)}$.
By the Chinese remainder theorem, solutions exist iff $a\equiv b\pmod{g}$ where $g=\gcd(\tau_l(0),\tau_s(0))$.
Setting $a = b+1$, if $g>1$, then $1\not\equiv 0\pmod{g}$ so no solutions exist.
If $g=1$, the map $t\mapsto(\tau_l(t),\tau_s(t))$ bijects $\mathbb{Z}/T_{\mathrm{eval}}\mathbb{Z}$ onto the full grid $\{1,\ldots,\tau_l(0)\}\times\{1,\ldots,\tau_s(0)\}$, so each pair $(b{+}1,b)$ occurs exactly once.
There are $\tau_s(0)$ such pairs.
For consecutivity: the transition $(b{+}1,b)\mapsto(b,b{-}1)$ preserves the gap-one condition for $b\ge 2$; only $(2,1)$ exits the set.
Hence all $\tau_s(0)$ occurrences form a single contiguous block.
\end{proof}

\begin{theorem}[Two-provider multi-period bound]
\label{thm:two_provider_multi}
For two providers with $\tau_l(0)>\tau_s(0)\ge 2$, a constrained adversary can force at most one additional infeasible match against CFM-SP compared to GSM over each hyper-period $T_{\mathrm{eval}}=\operatorname{lcm}(\tau_l(0),\tau_s(0))$.
If $\gcd(\tau_l(0),\tau_s(0))\ge 2$, the adversary cannot force any additional infeasible matches against CFM-SP.
The time-averaged regret differential is therefore at most $1/T_{\mathrm{eval}}$, which vanishes as capacities grow.
\end{theorem}

\begin{proof}
By Lemma~\ref{lem:sufficient_conditions}, the only regime where CFM-SP is vulnerable but GSM is not is the anti-sorted gap-one window $\tau_l(t)-\tau_s(t)=1$.
By Proposition~\ref{prop:algebraic_structure}, this window either does not exist~(when $\gcd>1$) or occurs as a single contiguous block of length~$\tau_s(0)$ per hyper-period~(when $\gcd=1$).
By Lemma~\ref{lem:one_per_period}, at most one infeasible match can be enforced within any single restake cycle, and the gap-one block falls within one restake cycle of the short provider.
Hence the adversary gains at most one extra infeasible match per hyper-period.
The time average is $1/T_{\mathrm{eval}} = 1/\operatorname{lcm}(\tau_l(0),\tau_s(0))$.
\end{proof}

\paragraph{Extension to $m$ providers.}
A natural question is whether the two-provider bound extends to $m$ providers.
To tackle this we might introduce a \emph{peel-and-load} characterization, wherein the adversary \emph{peels} the longest providers by sending short feasible jobs~(saving budget) then \emph{loads} shorter providers with infeasible jobs using the saved budget.
Under CFM-SP each peel saves $\tau_j - 1$ units~(a $1$-period job suffices to occupy the cheapest provider), whereas under GSM each peel saves only $\tau_j - \tau_{j+1} - 1$ units~(the adversary must send a job of length $\tau_{j+1}+1$ to occupy the shortest feasible provider).
The gap between these savings bounds the excess infeasible matches CFM-SP incurs.
Whether the two-provider $O(1)$-per-hyper-period bound decomposes additively across interacting provider pairs under aggregate workload constraints---and whether such a decomposition is tight---remains an important open question.

\section{Incomplete Information Extension}\label{app:incomplete_info}

CFM-SP resolves the market-level incentive problem; providers truthfully report their cost and availability, and the pricing layer clears the access margin.
However, once a provider is matched to a job, a separate execution-level problem arises.
The provider commits to~$w_d$ periods of compute at payment~$\theta$ per period, but the job may finish in fewer periods.
In practice, this imprecise job-length reporting is reasonable to expect either because the user conservatively overestimated $w_d$~(a rational strategy given that checkpointing makes underestimation recoverable at the, in practice, non-zero cost of a restart) or because the provider's hardware completes the work faster than the tier-minimum rate.
Under the base payment rule the provider collects $\theta\cdot w_d$ regardless of actual utilization, creating a moral hazard---providers who finish early have no incentive to report it and may idle for the remaining periods.
This appendix sketches a \emph{racing mechanism} that addresses this problem by having providers compete on completion time, with stakes that discipline the quotes.
The mechanism selects the fastest available provider in dominant strategies, reducing total idle time across the market; the payment structure itself remains unchanged, and eliminating idle-time rents entirely would require linking payment to quoted completion time~(discussed at the end of this section).

\paragraph{Setup.}
Fix a period~$t$ and a job~$d$ with duration~$w_d$.
Let $S_n\subseteq\mathcal{S}^t$ be the $n$ cheapest feasible providers selected by CFM~(Definition~\ref{def:cfm}).
Each provider~$s_i\in S_n$ privately observes a signal about their true completion time~$t_{s_i}(w_d)$ and forms an estimate~$\hat{t}_{s_i}$.
No provider's signal is disclosed to others.

\paragraph{Racing mechanism.}
Fix a tolerance $\varepsilon\ge 0$.
Consistent with Assumption~\ref{assump:verification_penalty}, each provider locks a stake~$L_{s_i}$ that is independent of their quote and strictly exceeds the maximum possible task reward,
\begin{equation}\label{eq:racing_stake}
L_{s_i} \;>\; P^t \cdot w_d\,.
\end{equation}
The mechanism then proceeds as follows: (i)~each $s_i\in S_n$ submits a completion-time quote~$\hat{t}_{s_i}$; (ii)~the job is awarded to $i^* \in \arg\min_{i\in S_n}\hat{t}_{s_i}$ at the posted price~$P^t$; (iii)~completion is deemed on-time iff $|t_{s_{i^*}}(w_d) - \hat{t}_{s_{i^*}}| \le \varepsilon$; if on-time, the stake is returned; otherwise the entire stake~$L_{s_i}$ is forfeited.

Since the stake is set before quotes are submitted and depends only on the posted price and job duration, providers cannot reduce their exposure by manipulating their quote.

\begin{proposition}[$\varepsilon$-accurate selection]
\label{prop:racing}
Under the racing mechanism with~\eqref{eq:racing_stake} and risk-neutral providers, every weakly undominated strategy for provider~$s_i$ satisfies
\[
\hat{t}_{s_i}
\;\in\;
\bigl[\,t_{s_i}(w_d)-\varepsilon,\;
t_{s_i}(w_d)+\varepsilon\,\bigr].
\]
Consequently, the mechanism selects a provider whose true completion time is within~$\varepsilon$ of the fastest among the $n$~racers, and the winning quote satisfies $\hat{t}_{i^*}\in [t^*(w_d)-\varepsilon,\,t^*(w_d)+\varepsilon]$, where $t^*(w_d):=\min_{i\in S_n}t_{s_i}(w_d)$.
\end{proposition}

\begin{proof}
If $s_i$ reports $\hat{t}_{s_i} < t_{s_i}(w_d) - \varepsilon$ and wins, the actual completion time exceeds the tolerance window with certainty, forfeiting the stake~$L_{s_i} > P^t\cdot w_d$.
Since the maximum reward from the job is $P^t\cdot w_d < L_{s_i}$, the net payoff is strictly negative.
Raising the report to $t_{s_i}(w_d)-\varepsilon$ weakly increases expected utility against any opponents' bids.

Conversely, if $s_i$ reports $\hat{t}_{s_i} > t_{s_i}(w_d)+\varepsilon$, lowering the report to $t_{s_i}(w_d)+\varepsilon$ weakly increases the probability of winning without affecting feasibility or payoff conditional on winning~(the stake and payment are independent of the quote).

Hence reports outside $[t_{s_i}(w_d)-\varepsilon,\,t_{s_i}(w_d)+\varepsilon]$ are weakly dominated.
Since each provider's undominated report preserves the ordering of true completion times, the winner has $t_{i^*}(w_d)=\min_{i\in S_n}t_{s_i}(w_d)$, and the bound on the winning quote follows.
\end{proof}

The racing mechanism integrates with CFM-SP as a refinement of the allocation stage. 
CFM first selects the $n$ cheapest feasible providers by reported cost, then the race selects the fastest among them by eliciting completion-time quotes.
Under the mechanism as stated, the winner is paid at the CFM-SP rate for the full job duration~$w_d$; the efficiency gain is that faster providers are matched first, reducing total idle time across the market.
A more aggressive variant may link payment to the quoted completion time~(paying $P^t\cdot\hat{t}_{s_{i^*}}$ rather than $P^t\cdot w_d$), which would eliminate idle-time rents entirely but would require an equilibrium analysis in place of the dominant-strategy guarantee of Proposition~\ref{prop:racing}---an interesting direction for future work.
The tolerance~$\varepsilon$ handles practical estimation noise in providers' self-assessment of their hardware speed; choosing~(or learning) $\varepsilon$ optimally as a function of the noise distribution is a further open question.

\section{Complexity Analysis and Pseudocode}\label{app:complexity}

We summarize the per-job time and space complexity of GSM~(Definition~\ref{def:GSM}) and CFM~(Definition~\ref{def:cfm}), then give pseudocode for each.
Throughout, $m=|\mathcal{S}^t|$ denotes the number of providers and $U$ denotes the number of distinct availability values.
For notational simplicity we write $\tau_i$ and $c_i$ for reported availability and cost~(dropping hats), since the complexity analysis is independent of whether reports are truthful.

\paragraph{Data structures and complexity.}
GSM requires a single lookup per arriving job, i.e.~find the idle provider with the shortest availability that exceeds the job duration.
A balanced binary search tree~(BST) keyed by availability supports this in $O(\log m)$ time per job, with $O(m)$ space.

CFM requires a \emph{two-key} lookup, i.e.~among all idle providers whose availability exceeds the job duration, find the one with minimum cost.
A segment tree over the $U$ distinct availability values, with each leaf storing a min-heap of $(c_i,\,i)$ pairs, supports this via a range-minimum query in $O(\log U + \log k_\tau)$ time per job, where $k_\tau$ is the maximum number of providers sharing a single availability value.
In the worst case this is $O(\log m)$. 
Space is $O(m + U)$.

\begin{table}[h]
\centering
\small
\renewcommand{\arraystretch}{1.15}
\begin{tabular}{lccc}
\toprule
\textbf{Algorithm} & \textbf{Data structure} &
\textbf{Per-job time} & \textbf{Space} \\
\midrule
GSM & Balanced BST on $\tau$
  & $O(\log m)$ & $O(m)$ \\
CFM & Segment tree + min-heaps
  & $O(\log m)$ & $O(m{+}U)$ \\
\bottomrule
\end{tabular}
\caption{Per-job complexity for GSM and CFM.
Both algorithms process $n$ jobs in $O(n\log m)$ total time.}
\label{tab:complexity}
\end{table}

\paragraph{Pseudocode: Greedy Shortest Matching.}

\begin{algorithmic}[1]
\State \textbf{State:} Balanced BST $T$ keyed by~$\tau$; each key
  stores a queue of provider IDs.
\Statex
\Function{Match}{$w_d$}
  \State $\mathrm{it} \gets
    T.\textsc{LowerBound}(w_d)$
    \Comment{first key with $\tau\ge w_d$}
  \If{$\mathrm{it} = T.\textsc{End}()$}
    \State \Return \textsc{Unmatched}
    \Comment{no feasible provider}
  \EndIf
  \State $\tau^* \gets \textsc{Key}(\mathrm{it})$
  \State $i^* \gets$ pop front of queue at $T[\tau^*]$
  \If{queue at $T[\tau^*]$ is empty}
    \State erase key $\tau^*$ from $T$
  \EndIf
  \State \Return $i^*$
\EndFunction
\end{algorithmic}

\medskip

\paragraph{Pseudocode: Cheapest-Feasible Matching.}

\begin{algorithmic}[1]
\State \textbf{State:} Segment tree over distinct availability
  values $\tau^{(1)}<\cdots<\tau^{(U)}$.
\Statex Each leaf $u$ holds a min-heap $H_u$ of $(c_i,\,i)$.
\Statex Each internal node stores $\textsc{Best}=(c,i,u)$: the
  minimum-cost provider in its subtree.
\Statex
\Function{Match}{$w_d$}
  \State $t \gets$ index such that
    $\tau^{(t)} < w_d \le \tau^{(t+1)}$
    \Comment{binary search}
  \State $(c,i^*,u^*) \gets
    \textsc{RangeMin}(t{+}1,\ldots,U)$
    \Comment{cheapest in suffix}
  \If{$c = +\infty$}
    \State \Return \textsc{Unmatched}
    \Comment{no feasible provider}
  \EndIf
  \State pop top of $H_{u^*}$
  \State \Call{UpdatePath}{$u^*$}
    \Comment{propagate from leaf to root}
  \State \Return $i^*$
\EndFunction
\end{algorithmic}

\medskip

In both algorithms, if no feasible provider exists for job~$d$~(i.e., every provider with $\tau_i\ge w_d$ is already assigned), the job is left unmatched; this is consistent with Definitions~\ref{def:GSM} and~\ref{def:cfm} in the main text.
The second-price payment~\eqref{eq:critical_value} is computed alongside the match. 
For CFM, the critical value is obtained by a second range-minimum query over $F_d\setminus\{i^*\}$, adding one additional $O(\log m)$ operation per matched job.
\end{document}